\documentclass[pre,10pt,twocolumn,
 amsmath,amssymb,
 aps,
]{revtex4-2}

\usepackage{graphicx}
\usepackage{dcolumn}
\usepackage{bm}
\usepackage[normalem]{ulem}

\begin{document}


\title{Stop-and-go locomotion of superwalking droplets}

\author{Rahil N. Valani}
\affiliation{School of Physics and Astronomy, Monash University, Victoria 3800, Australia}
 \author{Anja C. Slim}
\affiliation{School of Mathematics, Monash University, Victoria 3800, Australia}
\affiliation{School of Earth, Atmosphere and Environment, Monash University, Victoria 3800, Australia}
\author{Tapio P. Simula}%
\affiliation{Optical Sciences Centre, Swinburne University of Technology, Melbourne 3122, Australia}


\newcommand{\otol}{$(1,2,1)^{\text{L}}$}

\date{\today}

\begin{abstract}
Vertically vibrating a liquid bath at two frequencies, $f$ and $f/2$, having a relative phase difference $\Delta\phi_0$ can give rise to self-propelled superwalking droplets on the liquid surface. We have numerically investigated such superwalking droplets with the two driving frequencies slightly detuned, resulting in the phase difference $\Delta\phi(t)$ varying linearly with time. We predict the emergence of stop-and-go motion of droplets, consistent with  experimental observations [Valani et~al. Phys.~Rev.~Lett. {\bf 123}, 024503 (2019)]. Our simulations in the parameter space spanned by the droplet size and the rate of traversal of the phase difference uncover three different types of droplet motion: back-and-forth, forth-and-forth, and irregular stop-and-go motion. Our findings lay a foundation for further studies of dynamically driven droplets, whereby the droplet's motion may be guided by engineering arbitrary time-dependent functions $\Delta\phi(t)$.

\end{abstract}

\maketitle


\section{Introduction}

Intermittent locomotion where organisms alternate between active propulsion and a passive phase is frequently encountered in the natural world~\citep{interlocokramer,Gleiss2011}. It is observed in terrestrial, aquatic and aerial modes of locomotion in a diversity of species ranging from unicellular organisms such as ciliates to various reptiles, birds and mammals. Terrestrial organisms that exhibit intermittent locomotion typically come to a complete stop during the passive phase, but organisms in air or water may continue to glide forward~\citep{interlocokramer}. Intermittent locomotion has also been identified in artificial systems. For example, in artificial microswimmers that are powered by chemical activity, the self-interaction of the swimmer with its long-lived chemical wake results in speed bursts~\citep{stopgoswim}. In this paper, we explore the intermittent locomotion that emerges in a system of a self-propelled droplets on a vibrated liquid bath.

Vertically vibrating a bath of silicone oil with frequency $f$ can give rise to steadily walking droplets on the free surface of the liquid~\citep{Couder2005,Couder2005WalkingDroplets}. The walking droplet, also known as a walker, emerges just below the Faraday instability threshold above which the fluid-air interface becomes unstable to standing subharmonic Faraday waves of frequency $f/2$~\citep{Faraday1831a,Molacek2013DropsTheory}. Each bounce of a walker generates a localized, slowly decaying standing wave. The droplet interacts with these self-generated waves on subsequent bounces, resulting in a self-propelled droplet-wave entity on the liquid surface. Such walkers have been shown to mimic several peculiar behaviors that were previously though to be exclusive to the quantum realm~\citep{Bush2015,Bush2020review,Fort17515,harris_bush_2014,Oza2014,Perrard2014b,Perrard2014a,labousse2016,Zeeman,spinstates2018,PhysRevE.88.011001,Giletconfined2016,Saenz2017,Cristea,durey_milewski_wang_2020,Friedal,Eddi2009,tunnelingnachbin,tunneling2020,ValaniHOM,correlationnachbin,Durey2020hqft}. Recently, a new class of walking droplets, coined superwalkers, has been shown to emerge when the bath is vibrated simultaneously with two frequencies, $f$ and $f/2$, along with a relative phase difference $\Delta\phi_0$ between them~\citep{superwalker}. Superwalkers are typically bigger and faster than single-frequency driven walkers and their inter-droplet interactions give rise to novel multi-droplet behaviors.  

A given sized walker at a fixed driving amplitude and frequency typically has a fixed walking speed. By contrast, a given sized superwalker at fixed driving amplitudes and frequencies can have a range of walking speeds, dictated by the phase difference $\Delta\phi_0$. Depending on the value of $\Delta\phi_0$, small- to moderate-sized superwalkers ($0.4-0.7$\,mm in radius) can either be in the superwalking or pure bouncing regime while larger superwalkers (bigger than $0.7$\,mm in radius) can also coalesce with the bath for a narrow range of $\Delta\phi_0$ values. \citet{superwalker} observed that a slight detuning of the two driving frequencies to $f$ and $f/2+\epsilon$, results in a \emph{stop-and-go motion} (SGM) for superwalking droplets. In this locomotion, a droplet was observed to periodically switch between active self-propulsion and passive bouncing, see Fig.~\ref{fig: schematic}(a). In this paper, we adopt the theoretical model for superwalkers developed by \citet{superwalkernumerical} and explore the SGM of superwalking droplets through numerical simulations. We start by presenting the theoretical model in Sec.~\ref{theory} followed by a description of the emergence of SGM in Sec.~\ref{emergence}. In Sec.~\ref{PS}, we provide a description of various kinds of SGM observed in the parameter space spanned by the droplet size and the detuning parameter $\epsilon$ and explore them in detail in Sections~\ref{BF}, \ref{FF} and \ref{small detun}. We provide concluding remarks in Sec.~\ref{DC}.

\section{Theoretical formulation}\label{theory}

\begin{figure}
\centering
\includegraphics[width=\columnwidth]{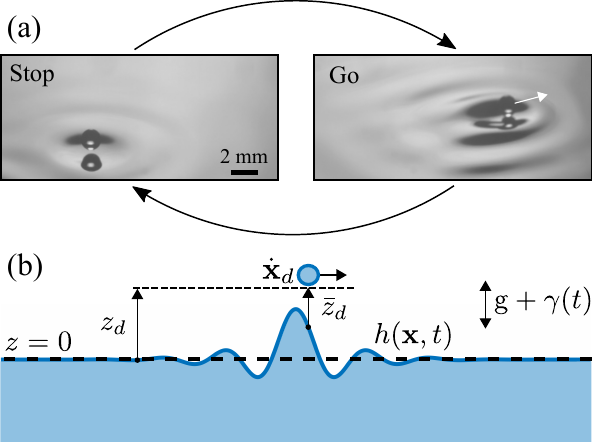}
\caption{Schematic of the droplet system. (a) Snapshots of a superwalker undergoing stop-and-go motion (SGM) showing the passive bouncing (stop) phase and the active self-propulsion (go) phase as observed in experiments. (b) The theoretical setup has a droplet located at a vertical position $z_d$ with an underlying wave field $h(\mathbf{x},t)$ and walking horizontally with velocity $\dot{x}_d$ in the frame of reference of the driven bath. The notation is defined in the main text. The vertical amplitude of the wave field is exaggerated.}
\label{fig: schematic}
\end{figure}

Consider a liquid droplet of mass $m$ and radius $R$ walking on a bath of the same liquid of density $\rho$, viscosity $\nu$ and surface tension $\sigma$, as shown schematically in Fig.~\ref{fig: schematic}(b). The bath is subjected to vertical time varying acceleration $\gamma(t)$. The system geometry is described in the co-moving frame of the bath by horizontal coordinates $\mathbf{x} = (x,y)$ and the vertical coordinate $z$, with the origin chosen to be the undeformed surface of the bath (dashed horizontal line). In the bath's frame of reference, the center of mass of the droplet is located at a horizontal position $\mathbf{x}_d$ and the south pole of the droplet at a vertical position $z_d$ such that $z_d=0$ represents initiation of the droplet's impact with the undeformed surface of the bath. The free surface elevation of the liquid filling the bath is at $z=h(\mathbf{x},t)$.

\citet{superwalkernumerical} developed a theoretical model for superwalkers by considering two-frequency driving of the form $\gamma(t)=\gamma_f\sin(2\pi f t)+\gamma_{f/2}\sin(\pi f t+\Delta\phi_0)$, where $\gamma_f$ is the amplitude of the primary driving frequency, $\gamma_{f/2}$ is the amplitude of the subharmonic frequency and $\Delta\phi_0$ is the constant relative phase difference between the two. In the experiments of \citet{superwalker}, SGM was observed when the bath was driven at frequencies $f$ and $f/2+\epsilon$ giving rise to time varying acceleration: $\gamma_f\sin(2\pi f t)+\gamma_{f/2+\epsilon}\sin(\pi f t+\Delta\phi_0+2\pi\epsilon t)$. 
This can be rewritten as
\begin{equation}\label{drivingform}
\gamma(t)=\gamma_f\sin(2\pi f t)+\gamma_{f/2}\sin(\pi f t+\Delta\phi(t)),
\end{equation}
where $\Delta\phi(t)=\Delta\phi_0+2\pi\epsilon t$, and $\gamma_{f/2} \approx \gamma_{f/2+\epsilon}$ for $\epsilon\ll f/2$. Hence, two frequency driving at $f$ and $f/2+\epsilon$ can be interpreted as driving at $f$ and $f/2$ with a continuously changing phase difference $\Delta\phi(t)$ that varies linearly with time. Assuming that $\epsilon\ll f/2$ so that $\Delta\phi(t)$ evolves slowly, we can use a `quasi-static' approximation and use the model developed by \citet{superwalkernumerical} for a constant phase difference $\Delta\phi_0$ and apply it to the case of time varying phase difference $\Delta\phi(t)$. Hence, we will use the driving acceleration prescribed by Eq.~\eqref{drivingform} in the model of \citet{superwalkernumerical} to study the SGM. We proceed by reviewing the key equations governing the vertical dynamics, horizontal dynamics and wave form in the superwalker model of \citet{superwalkernumerical}.

\subsection{Vertical dynamics}
The vertical motion of the droplet is governed by
\begin{equation}\label{eq: vertical}
    m\ddot{z}_d=-m[\text{g}+\gamma(t)]+{F_N}(t).
\end{equation}
In this equation, the first term on the right hand side is the effective gravitational force acting on the droplet in the oscillating frame of the bath, with g the constant acceleration due to gravity. The second term on the right hand side is the normal force imparted to the droplet during contact with the liquid surface. This contact force is calculated by modeling the bath as a spring and a damper~\cite{molacek_bush_2013}, 
\begin{equation}
{F_N}(t)=H(-\bar{z}_d) \,\max\left(-k\bar{z}_d-b\dot{\bar{z}}_d,0\right).
\label{springdamp}
\end{equation}
In Eq.~(\ref{springdamp}), $H(\cdot)$ stands for the Heaviside step function and $\bar{z}_d=z_d-h(\mathbf{x}_d,t)$ is the vertical position of the droplet above the free surface of the bath. The constants $k$ and $b$ are the spring constant and damping force coefficient, respectively.

\subsection{Wave field}
The free surface elevation $z=h(\mathbf{x},t)$ is calculated by adding individual waves generated by the droplet on each bounce:
\begin{equation}\label{eq: wave}
\textstyle h(\mathbf{x},t)=\sum_{n} h_n(\mathbf{x},\mathbf{x}_n,t,t_n)\, ,
\end{equation}
where $h_n(\mathbf{x},\mathbf{x}_n,t,t_n)$ is the wave field generated by bounce $n$ at location $\mathbf{x}_n$ and time $t_n$. 
The individual waves generated by the droplet on each bounce are localized decaying Faraday waves, which for the case of the bath being driven at frequencies $f$ and $f/2$ take the form
\begin{widetext}
\begin{align}\label{sw_wave}
h_n(\mathbf{x},t)= A_{f/2}\frac{\cos(\pi f t +\theta_{F/2}^{+})}{\sqrt{t-t_n}} \,\text{J}_0(k_{F/2}|\mathbf{x}-{\mathbf{x}}_n|)\,\exp{\left[ -\frac{t-{t}_n}{T_F \text{Me}_{f/2}} - \frac{T_F |\mathbf{x}-{\mathbf{x}}_n|^2}{8 \pi D_{f/2} (t-{t}_n)} \right]}& \nonumber\\
+ A_{f/4}\frac{\cos(\pi f t/2 + \theta_{F/4}^{+})}{\sqrt{t-t_n}} \,\text{J}_0(k_{F/4}|\mathbf{x}-{\mathbf{x}}_n|)\,\exp{\left[ -\frac{t-{t}_n}{T_F\text{Me}_{f/4}} - \frac{T_F |\mathbf{x}-{\mathbf{x}}_n|^2}{8 \pi D_{f/4} (t-t_n)} \right]}&,
\end{align}
\end{widetext}
where the impact location $\mathbf{x}_n$ and the time of impact $t_n$ are given by
\begin{equation*}
    \mathbf{x}_n=\int_{t_n^i}^{t_n^c}\mathbf{x}_d(t')F_N(t')\,\text{d}t'\Big{/}\int_{t_n^i}^{t_n^c}F_N(t')\,\text{d}t',
\end{equation*}
and
\begin{equation*}
    t_n=\int_{t_n^i}^{t_n^c}t'F_N(t')\,\text{d}t'\Big{/}\int_{t_n^i}^{t_n^c}F_N(t')\,\text{d}t',
\end{equation*}
%
where $t_n^i$ and $t_n^c$ are the time of initiation and completion of the $n$th impact. The interpretation of Eq.~(\ref{sw_wave}) is that a droplet bouncing under the prescribed two-frequency driving excites two dominant subharmonic standing waves with amplitude coefficients $A_{f/2}$ and $A_{f/4}$ and wavenumbers $k_{F/2}$ and $k_{F/4}$, corresponding to  frequencies of $f/2$ and $f/4$. These waves decay in time with a time scale $T_F \text{Me}_{f/2}$ and $T_F \text{Me}_{f/4}$, where $T_F=2/f$ is the period of $f/2$ Faraday waves. The spatial structure of each component of the wave takes the form of a Bessel function of the first kind and zeroth order, $\text{J}_0(\cdot)$.
The waves also spread diffusively with diffusion coefficients $D_{f/2}$ and $D_{f/4}$, and have phase shifts $\theta_{F/2}^{+}$ and $\theta_{F/4}^{+}$ with respect to the driving signal. Due to the evolving phase difference $\Delta\phi(t)$, the phase shift $\theta_{F/4}^{+}(t)$ will vary with time. We refer the reader to \citet{superwalkernumerical} for explicit equations for these parameters. We note that although we have considered both $f/2$ and $f/4$ waves in the wave field of the superwalker, the dominant contribution to the wave field comes from the $f/2$ waves in the parameter regime where superwalking is realized~\citep{superwalkernumerical}.

\subsection{Horizontal dynamics}

The horizontal dynamics of the droplet is modeled using the following equation~\citep{Molacek2013DropsTheory}:
\begin{equation}\label{eq: hor}
    m\ddot{\mathbf{x}}_d+D_{tot}(t)\dot{\mathbf{x}}_d=-{F_N}(t) \boldsymbol{\nabla} h(\mathbf{x}_d,t),
\end{equation}
where $D_{tot}(t)=D_{mom}(t)+D_{air}$. Here, $D_{mom}(t)=C\sqrt{\frac{\rho R}{\sigma}}{F_N}(t)$ is the drag comprising of momentum loss during contact with the bath and $D_{air}=6\pi R \mu_a$ is the air drag. Here $C$ is the contact drag coefficient and $\mu_a$ is the dynamic viscosity of air. The force on the right hand side is the horizontal component of the contact force arising from the small slope $|\nabla h(\mathbf{x}_d,t)|\ll 1$ of the underlying wave field.

\subsection{Numerical implementation}

We restrict the simulations to one horizontal dimension since even if two-dimensional planar dynamics were implemented, the SGM still proceeds along a line. 
Eqs.~\eqref{eq: vertical} and \eqref{eq: hor}  corresponding to vertical and horizontal equations of motions respectively are solved using the leap-frog method~\citep{Sprott}, a modified version of the Euler method where the new horizontal and vertical positions of the droplet are calculated using the old velocities and then the new velocities are calculated using the new positions. Converting the second order differential equation for the vertical dynamics in Eq.~(\ref{eq: vertical}) into a system of two first order ordinary differential equations and discretizing using the leap-frog method yields,
\begin{equation*}\label{eq: Vertical disc 1}
    z_d(t_{i+1})=z_d(t_{i})+\Delta t\,v_d(t_{i}),
\end{equation*}
and
\begin{equation*}\label{eq: Vertical disc 2}
    v_d(t_{i+1})=v_d(t_{i})+\frac{\Delta t}{m} \left[-m(\text{g}+\gamma(t_{i+1}))+F_N(t_{i+1})\right],
\end{equation*}
where $v_d(t)=\dot{z}_d(t)$ and
\begin{equation*}\label{eq: Vertical disc 3}
{F_N}(t_{i+1})=H(-\bar{z}_d(t_{i+1})) \,\max\left(-k\bar{z}_d(t_{i+1})-b\bar{v}_d(t_{i}),0\right).
\end{equation*}
Here $\bar{z}_d(t_{i+1})=z_d(t_{i+1})-h(x_d(t_{i+1}),t_{i+1})$ and $\bar{v}_d(t_{i})=v_d(t_{i})-\frac{\partial h}{\partial t}(x_d(t_{i+1}),t_{i+1})$. The total wave height beneath the droplet $h(x_d(t_{i+1}),t_{i+1})$ is calculated using Eq.~\eqref{sw_wave} by keeping the waves from the last $N$ impacts of the droplet. For the integral required to calculate the location of impact $\mathbf{x}_n$, the time of impact $t_n$ and the amplitudes $A_{f/2}$ and $A_{f/4}$ we used the MATLAB inbuilt trapezoid function. 

Similarly, Eq.~(\ref{eq: hor}) governing the horizontal dynamics takes the following form,
\begin{equation*}\label{eq: Horizontal disc 1}
    x_d(t_{i+1})=x_d(t_{i})+\Delta t\,u_d(t_{i}),
\end{equation*}
and
\begin{align*}\label{eq: Horizontal disc 2}
    u_d(t_{i+1})=u_d(t_{i})+\frac{\Delta t}{m}\Big[-D_{tot}(t_{i+1})u_d(t_{i})&\nonumber\\
    -{F_N}(t_{i+1}) \frac{\partial h}{\partial x}(x_d(t_{i+1}),t_{i+1})\Big]&,
\end{align*}
where $u_d(t)=\dot{x}_d(t)$.

The physical parameters were fixed to match the experiments of \citet{superwalker}: $\rho=950$\,$\text{kg}/\text{m}^3$, $\nu=20$\,cSt and $\sigma=20.6$\,$\text{mN}/\text{m}$. We also 
fix the driving frequencies to $f=80$\,Hz and $f/2=40$\,Hz, and the acceleration amplitudes to $\gamma_{f}=3.8$\,g and $\gamma_{f/2}=0.6$\,g with an initial phase difference $\Delta\phi_0=0^{\circ}$. There are three adjustable parameters in the model: the spring constant of the bath $k$, the damping coefficient of the bath $b$ and the dimensionless contact drag coefficient $C$. The corresponding dimensionless parameters are given by $K=k/m\omega_d^2$ and $B=b/m\omega_d$, where $\omega_d=\sqrt{\sigma/\rho R^3}$ is the droplet's characteristic oscillation frequency~\cite{molacek_bush_2013}. We choose $C=0.17$, $K=0.70$ and $B=0.60$ as these values give a good fit to the experimental data for small- to moderate-sized superwalkers~\citep{superwalkernumerical}.

To increase computational speed, we only stored the waves generated by the $N=100$ most recent bounces of the droplet and discarded the earlier ones, which have typically decayed to below $10^{-5}$ of their initial amplitude for the chosen parameters. A time step of $\Delta t =T_F/100$ was used. The simulations were initialized with $x_d=0\,\text{mm}$, $u_d=1\,\text{mm/s}$, $v_d=0$\,mm/s and three different vertical positions $z_d= (0,5,10)R$. Multiple initial conditions were used so that different modes of walking and bouncing existing for the same parameter values are likely to be captured.

\section{Emergence of Stop-and-go motion (SGM)}\label{emergence}

\begin{figure*}
\centering
\includegraphics[width=2\columnwidth]{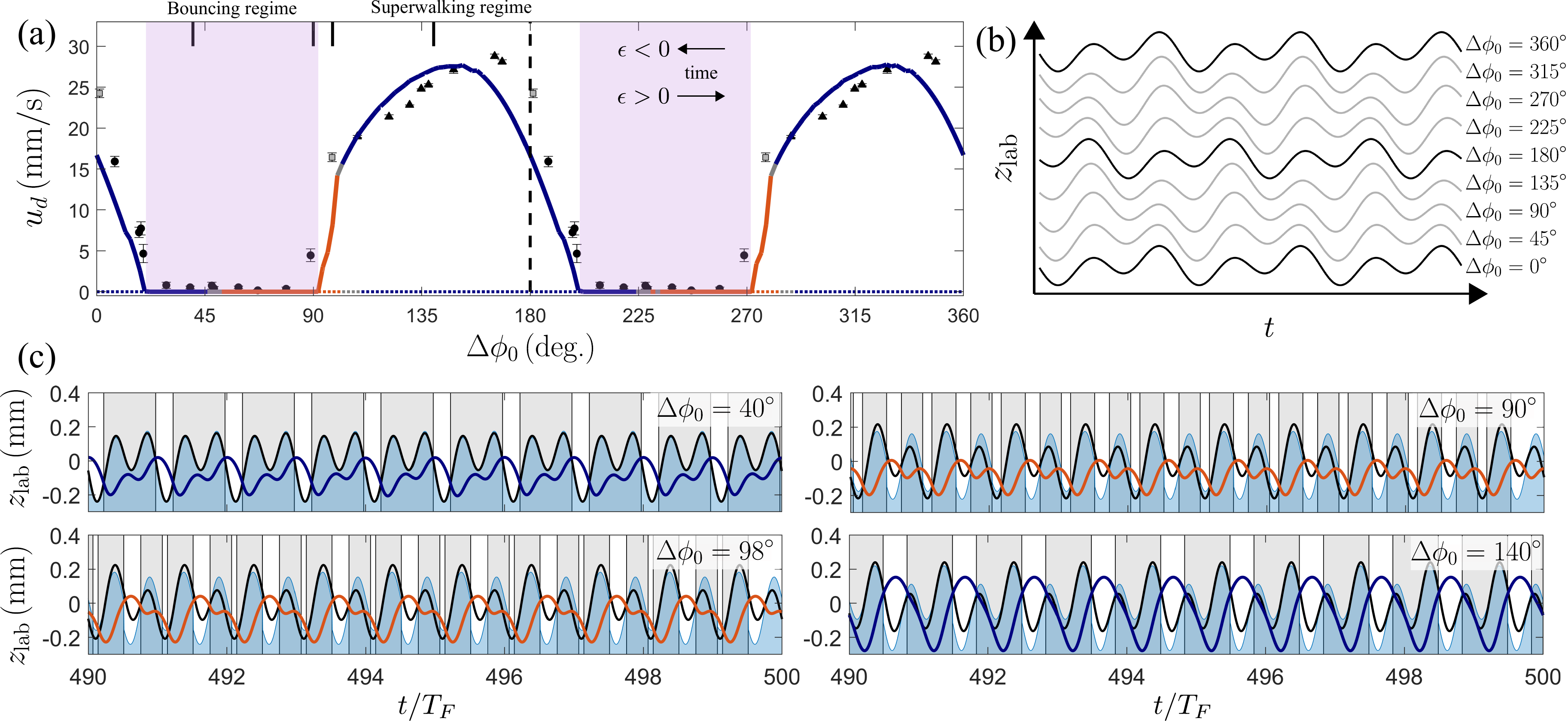}
\caption{Droplet dynamics as a function of the relative phase difference. (a) Steady walking speed $u_d$ of a superwalker as a function of a fixed phase difference $\Delta\phi_0$ for a droplet of radius $R=0.60$\,mm from experiments of \citet{superwalker} (markers) and simulations of \citet{superwalkernumerical} (multi-colored curve). The experimental data to the right of the vertical dashed line at $\Delta\phi_0=180^\circ$ is repeated. The style of marker indicates the bouncing modes observed in experiments: $(1,2,2)$ are black circles $\bullet$, $(1,2,1)$ are black triangles {\protect \scalebox{0.8}{$\blacktriangle$}}, transition between a $(1,2,1)$ and a $(1,2,2)$ mode are gray squares {\protect \scalebox{0.5}{$\blacksquare$}}. The colors on the solid curve indicate different vertical bouncing mode observed in simulations: \otol{} in navy blue, $(1,2,2)$ in red, and chaotic modes that arise at the transition between \otol{} and $(1,2,2)$ are shown in gray. The light purple region indicates the bouncing regime where a zero superwalking speed, in other words pure bouncing, is observed. The dotted multi-colored horizontal line in the superwalking regime indicates pure bouncing states that are unstable to horizontal perturbations. In this panel, a time-varying phase difference $\Delta\phi(t)=2\pi \epsilon t$ would correspond to traversing the $\Delta\phi_0$ axis from either left to right $(\epsilon>0)$ or right to left $(\epsilon<0)$. Panel (b) shows the vertical periodic motion of the bath, $\mathcal{B}(t)=-(\gamma_f/(2\pi f)^2)\sin(2\pi f t)-(\gamma_{f/2}/(\pi f)^2)\sin(\pi f t+\Delta\phi_0)$, in the lab frame at different constant phase differences $\Delta\phi_0$. Panel (c) shows vertical bouncing modes obtained for different values of $\Delta\phi_0$ that are marked by a black line segment at the top of panel (a). In this panel, the solid black curves indicate the bath motion, $\mathcal{B}(t)$, the multi-colored curves represent the vertical motion of the south pole of the droplet, $z_d(t)+\mathcal{B}(t)$, and the filled blue regions illustrate the vertical motion of the liquid surface directly beneath the droplet, $h(x_d,t)+\mathcal{B}(t)$, all in the lab frame. The gray regions indicate times at which the droplet is in contact with the bath.}
\label{Fig: stopgo1}
\end{figure*}

A given sized superwalker at fixed driving amplitudes and frequencies can have a range of walking speeds $u_d$ depending on the phase difference $\Delta\phi_0$. This walking speed dependence on the phase difference of superwalkers from experiments (markers) and numerical simulations (multi-colored curve) is shown for a typical droplet of radius $R=0.60$\,mm in Fig.~\ref{Fig: stopgo1}(a). We find two different regimes for such a droplet: a bouncing regime where the droplet bounces in place with no horizontal motion, and a superwalking regime where the droplet also self-propels horizontally while bouncing. However, we note that in the superwalking regime, there exist purely bouncing states (dotted multi-colored horizontal line in the superwalking regime) that are unstable to horizontal perturbations. The two regimes realized for $R=0.60$\,mm droplet is typical for small- to moderate-sized superwalkers~\citep{superwalker}. Note that due to the form of driving considered here at frequencies $f$ and $f/2$, the driving signal, see Fig.~\ref{Fig: stopgo1}(b), and hence the walking speed dependence of the phase difference in the region $180^{\circ}\leq \Delta\phi_0 \leq 360^{\circ}$ is a repeat from the region $0^{\circ}\leq \Delta\phi_0 \leq 180^{\circ}$. 

To describe the vertical dynamics of the droplet, we follow \citet{superwalker} and use the notation $(l,m,n)$ to indicate that the droplet impacts the surface $n$ times during $m$ oscillation periods of the bath at frequency $f$, which equals $l$ oscillation periods of the bath at frequency $f/2$. In the superwalking regime, we find that the droplet is typically bouncing in a $(1,2,1)^\text{L}$ mode, where the superscript `L' denotes a low-bouncing, long-contact mode compared to a high-bouncing, short-contact mode that is also observed for smaller superwalkers~\citep{superwalkernumerical}. The $(1,2,1)$ bouncing mode is crucial for walking as the droplet in this mode is bouncing at the same frequency as the frequency of the subharmonic Faraday waves that emerge beyond the Faraday instability threshold. Thus, the droplet's bouncing is in resonance with the damped Faraday waves it generates and with which it interacts. In the bouncing regime and at the start of the superwalking regime, we also find a $(1,2,2)$ bouncing mode where the droplet contacts the bath twice, typically a high bounce and a low bounce, every two up-and-down cycles of the bath. The bouncing modes from simulations match well with experiments, however, for some values of $\Delta\phi_0$, we find \otol{} mode where a $(1,2,2)$ is observed in experiments. We note that it is difficult to distinguish between a \otol{} and a $(1,2,2)$ mode in experiments and hence it is not clear whether all the $(1,2,2)$ modes observed in experiments of \citet{superwalker} are truly $(1,2,2)$ or some in fact may even be \otol{}. The predicted vertical dynamics of the droplet for a selection of $\Delta\phi_0$ values is shown in Fig.~\ref{Fig: stopgo1}(c).
 
By allowing the phase difference $\Delta\phi_0$ to vary slowly according to $\Delta\phi(t)=2\pi\epsilon t$, we can continuously traverse the $\Delta\phi_0$ axis in Fig.~\ref{Fig: stopgo1}(a) either from left to right ($\epsilon>0$) or right to left ($\epsilon<0$). This gives rise to periodic traversals of the bouncing regime and the superwalking regime resulting in the emergence of SGM. The rate of traversal is controlled by $\epsilon$ with the period of one cycle of traversal, in other words one SGM cycle, given by $1/|\epsilon|$ corresponding to phases sweeping from $0^{\circ}$ to $360^{\circ}$. 

Once the phase difference is allowed to evolve with time, the system is subjected to three characteristic time scales set by the input parameters: (i) the time scale, $T_F$, of the bath driving or the oscillation of the underlying dominant $f/2$ standing waves generated by the droplet on each bounce, (ii) the memory time scale, $T_F \text{Me}_{f/2}$, of temporal decay of the dominant $f/2$ waves generated by the droplet and (iii) the SGM time scale, $1/|\epsilon|$, introduced by the evolving phase difference $\Delta\phi(t)$. The interactions of these time scales along with the time scale for the inertial response of the droplet, give rise to the emergent droplet dynamics where the vertical bouncing dynamics of the droplet occurs on the time scale (i), and the evolution of the horizontal walking dynamics of the droplet occurs on the time scale (iii). 

 \begin{figure*}
\centering
\includegraphics[width=2\columnwidth]{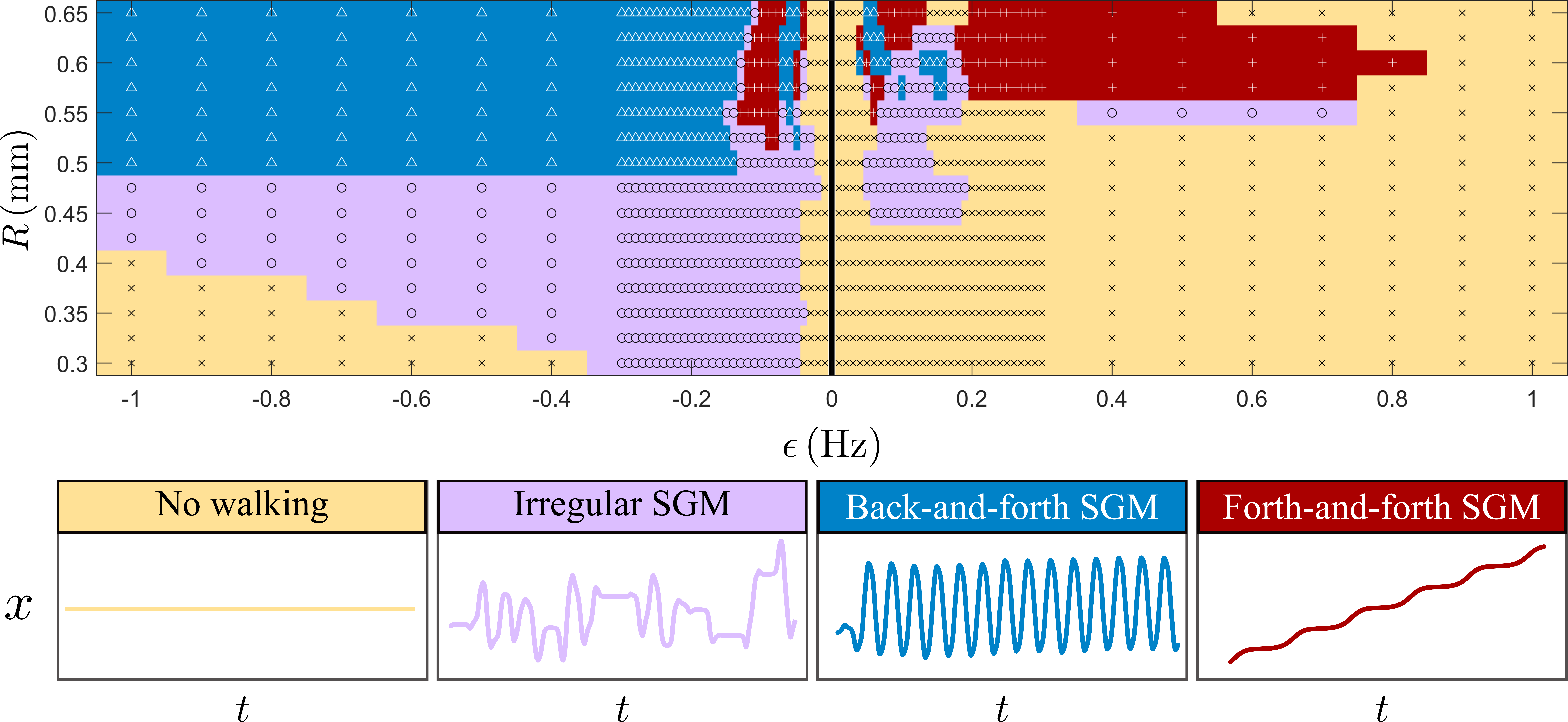}
\caption{Various types of SGM observed from simulations in the $(\epsilon,R)$ parameter space. Four distinct types of dynamics are observed: (i) No walking ($\times$, beige region) where the droplet bounces vertically but does not propel horizontally, (ii) irregular SGM ($\circ$, purple region) where the droplet erratically switches its walking direction and/or traverses unequal distances after every half SGM cycle, (iii) back-and-forth SGM ($\Delta$, blue region) where the droplet switches direction after every half SGM cycle and (iv) forth-and-forth SGM (+, maroon region) where the droplet moves in the same direction after every half SGM cycle. The markers indicate the parameter values where simulations were performed while the background colors represent the interpolated region.}
\label{Fig: stopgo_PS}
\end{figure*}

\section{Parameter-space description}\label{PS}

\begin{figure*}
\centering
\includegraphics[width=2\columnwidth]{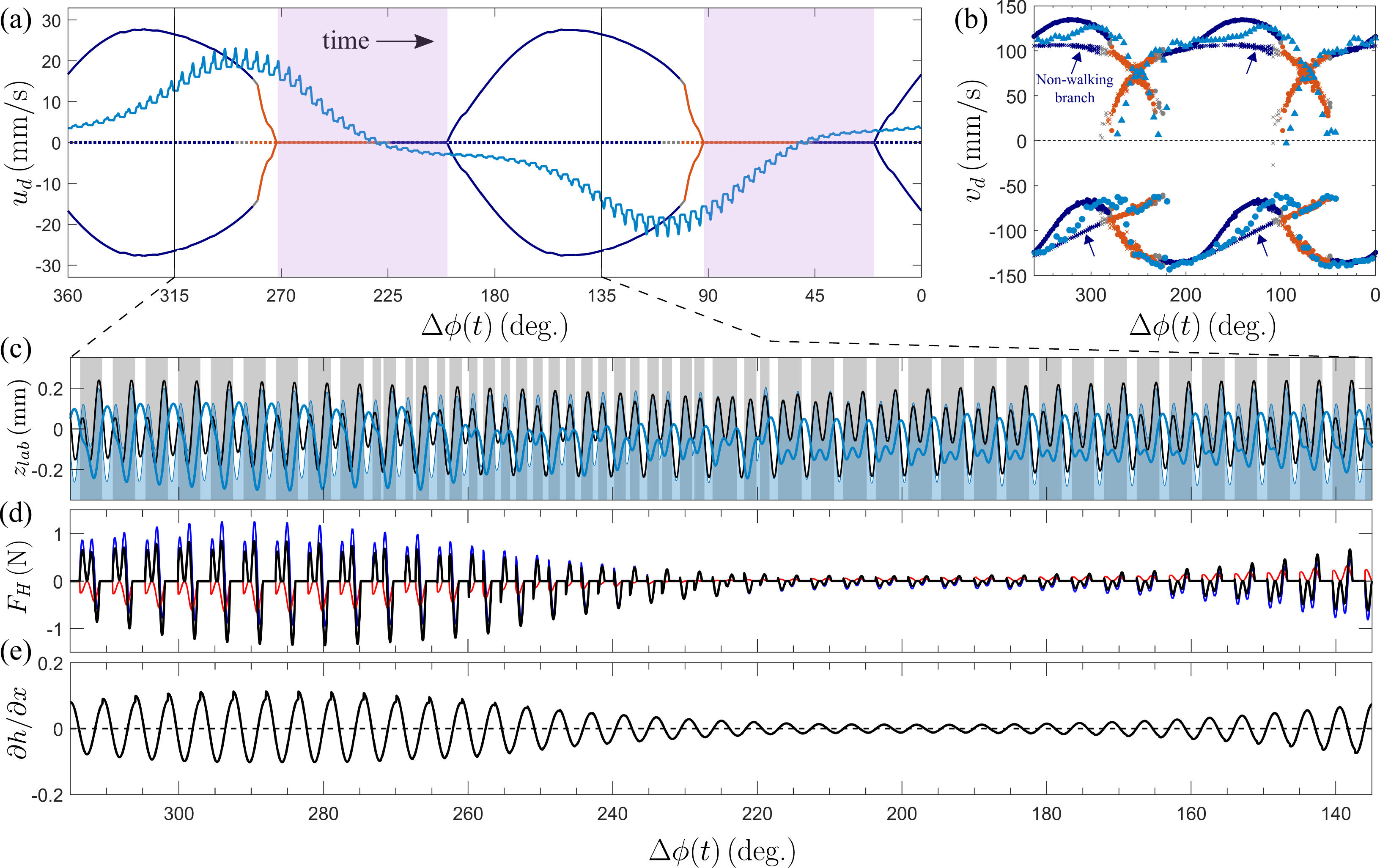}
\caption{Back-and-forth SGM: (a) Horizontal walking speed $u_d$ (blue curve) as a function of the evolving phase difference $\Delta\phi(t)=2\pi\epsilon t$ for one cycle of the SGM. The walking speed from constant phase difference superwalker simulations from Fig.~\ref{Fig: stopgo1}(a) is shown as a multi-colored curve. The light purple region indicates the pure bouncing regime. Panel (b) shows the vertical velocity $v_d$ of the droplet at impact (lower frame) and lift-off (upper frame) from constant phase superwalker simulations (background blue and red multi-colored markers with $\bullet$ indicating the superwalking solution branch and $\times$ indicating the non-walking, pure bouncing solution branch in the superwalking regime) and the back-and-forth SGM (blue $\bullet$ indicating impact and blue {\protect \scalebox{0.8}{$\blacktriangle$}} indicating lift-off). Panel (c) shows the vertical motion of the bath (black curve), wave height beneath the droplet (blue shaded region) and the vertical location of the droplet's `south pole' (blue curve) during back-and-forth SGM. The gray shaded region represents contact between the droplet and the underlying wave. Panel (d) shows the evolution of the horizontal wave force $F_{W}=-{F_N}(t)\partial h/\partial x$ (blue), horizontal drag force $F_{D}=-D_{tot}(t)\dot{x}_d$ (red) and the total horizontal force $F_{H}=F_{W}+F_{D}$ (black) during the back-and-forth SGM. Panel (e) shows the evolution of the gradient of the wave beneath the droplet $\partial h/\partial x$ (black solid curve) with the dashed line indicating zero gradient. The droplet size is fixed to $R=0.60$\,mm and the detuning is fixed to $\epsilon=-0.5$\,Hz.}
\label{Fig: backforth1}
\end{figure*}

We have simulated droplets in the parameter space formed by the detuning parameter $\epsilon$ that prescribed the rate at which the different phase differences are traversed and the droplet radius $R$. For the chosen parameters, the time scale for the bouncing dynamics or the oscillations of the waves generated by the droplet is $T_F=2/f=0.025$\,s, while the memory time scale is $T_F \text{Me}_{f/2} \approx 0.22$\,s. We explore the parameter space in the range $-1\leq \epsilon \leq 1$\,Hz and $0.30 \leq R \leq 0.65$\,mm. We have chosen $|\epsilon|\leq1$\,Hz to ensure $\epsilon\ll f/2$ so that our quasi-static approximation remains valid. The minimum droplet size is set by the start of the walking regime of droplets as droplets smaller than this are typically unable to walk at these parameters~\citep{superwalker,superwalkernumerical}. We restrict the maximum droplet size to $R=0.65$\,mm as the theoretical model used here for superwalkers fails to capture experimentally observed characteristics for larger droplets~\citep{superwalkernumerical}. The different dynamical behaviors observed in the parameter-space plot are shown in Fig.~\ref{Fig: stopgo_PS}. We have observed four qualitatively different types of dynamics in the $(\epsilon,R)$ parameter space: (i) no walking, (ii) irregular SGM, (iii) back-and-forth SGM and (iv) forth-and-forth SGM. In the irregular SGM, we find that the droplet erratically reverses its walking direction and/or travels a variable distance after every half cycle of the SGM. In the back-and-forth SGM, intriguingly, the droplet reverses its walking direction after every half cycle of the SGM resulting in the droplet oscillating back-and-forth about a fixed mean position or the droplet performs back-and-forth oscillations with a drift in the mean position. Conversely, in the forth-and-forth SGM, the droplet maintains the same walking direction after every half cycle of the SGM. 

The parameter-space diagram in Fig.~\ref{Fig: stopgo_PS} shows that relatively small droplets ($0.30\lesssim R \lesssim 0.50$\,mm) are either unable to walk under this prescribed driving or undergo irregular SGM for small negative detunings. Relatively larger droplets ($0.50\lesssim R \lesssim 0.65$\,mm) show a wide range of stop-and-go behavior including back-and-forth and forth-and-forth motion. For these we typically observe only back-and-forth SGM for relatively large magnitude negative detunings ($-1.0 \lesssim \epsilon \lesssim -0.2\,$Hz) and typically forth-and-forth SGM for relatively large magnitude positive detunings ($0.2 \lesssim \epsilon \lesssim 0.7\,$Hz). For both positive and negative detunings that are small in magnitude, we see a mix of back-and-forth, forth-and-forth and irregular SGM. Moreover, for very small magnitudes of detunings $|\epsilon|\leq0.04$\,Hz, we find that the SGM ceases. To understand these different behaviors in more detail, we study a fixed droplet size of $R=0.60$\,mm, which shows a diversity of behaviors. We start by exploring the back-and-forth SGM in Sec.~\ref{BF}, followed by forth-and-forth SGM in Sec.~\ref{FF} and the small detuning regime in Sec.~\ref{small detun}.

\begin{figure*}
\centering
\includegraphics[width=2\columnwidth]{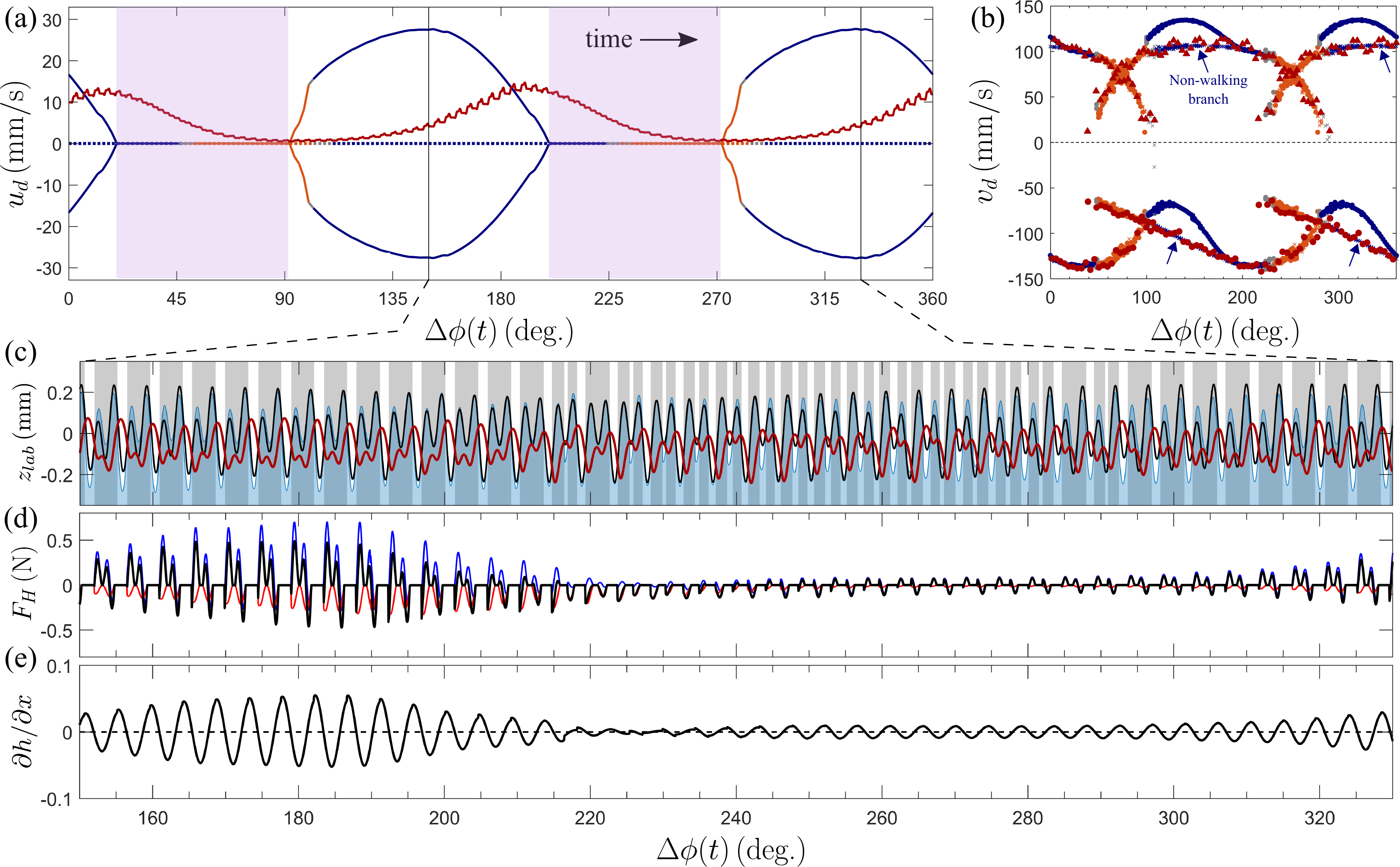}
\caption{Forth-and-forth SGM: (a) Horizontal walking speed $u_d$ (maroon curve) as a function of the evolving phase difference $\Delta\phi(t)=2\pi\epsilon t$ for one cycle of the SGM. The walking speed from constant phase difference superwalker simulations from Fig.~\ref{Fig: stopgo1}(a) is shown as a multi-colored curve. The light purple region indicates the bouncing regime. Panel (b) shows the vertical velocity $v_d$ of the droplet at impact (lower frame) and lift-off (upper frame) from constant phase superwalker simulations (background blue and red multi-colored markers with $\bullet$ indicating the superwalking solution branch and $\times$ indicating the non-walking, pure bouncing solution branch in the superwalking regime) and the forth-and-forth SGM (maroon $\bullet$ indicating impact and maroon {\protect \scalebox{0.8}{$\blacktriangle$}} indicating lift-off). Panel (c) shows the vertical motion of the bath (black curve), wave height beneath the droplet (blue shaded region) and the vertical location of the droplet's `south pole' (maroon curve) during forth-and-forth SGM. The gray shaded region represents contact between the droplet and the underlying wave. Panel (d) shows the evolution of the horizontal wave force $F_{W}=-{F_N}(t)\partial h/\partial x$ (blue), horizontal drag force $F_{D}=-D_{tot}(t)\dot{x}_d$ (red) and the total horizontal force $F_{H}=F_{W}+F_{D}$ (black) during the forth-and-forth SGM. Panel (e) shows the evolution of the gradient of the wave beneath the droplet $\partial h/\partial x$ (black solid curve) with the dashed line indicating zero gradient. The droplet size is fixed to $R=0.60$\,mm and the detuning is fixed to $\epsilon=0.5$\,Hz.}
\label{Fig: forthforth1}
\end{figure*}

\section{Back-and-forth SGM}\label{BF}

For relatively large droplets and typically negative detunings in Fig.~\ref{Fig: stopgo_PS}, we find that in the SGM, the droplet reverses its walking direction after every half SGM cycle resulting in a back-and-forth motion. One cycle of a typical back-and-forth SGM for $R=0.60$\,mm and $\epsilon=-0.5$\,Hz is shown in Fig.~\ref{Fig: backforth1} (see also Supplementary Material~\citep{supplement} for a video). During the acceleration period of this SGM, the droplet bounces in a $(1,2,1)^{\text{L}}$ mode where the droplet is efficiently able to generate a slowly decaying localized Faraday waves on each bounce and thus build up the wave field that propels it, see Fig.~\ref{Fig: backforth1}(c) and (e). In this period, during contact the droplet receives two horizontal impulses in the direction of its motion and one horizontal impulse in the opposite direction such that the overall horizontal force, sum of the force from the underlying wave field and the net drag force, is in the direction of the droplet's motion, see Fig.~\ref{Fig: backforth1}(d). 
Since the overall horizontal force is imparted in the direction of motion of the droplet, the droplet accelerates. Note from Fig.~\ref{Fig: backforth1}(a) that during this acceleration stage, the droplet's instantaneous velocity $u_d$ (blue curve) is lower than the corresponding steady superwalking velocity for a corresponding constant phase difference driving (multi-colored curve).

Once the instantaneous speed of the droplet exceeds the corresponding steady superwalking speed for a constant phase difference driving, further evolution in this bouncing mode results in an overall horizontal force in the direction opposite to $u_d$ causing the droplet to decelerate. 
This is soon followed by the droplet transitioning to a $(1,2,2)$ mode where the droplet impacts the bath twice in two up-and-down motions of the bath. In this mode, during one period of the bath, the droplet generates two waves of opposite phase that interfere destructively and thus their contribution to building up the overall wave field is small. This results in the decay of the previously built-up wave field. At this time in the deceleration stage, the droplet is in the bouncing regime and two competing effects are taking place simultaneously: (i) the horizontal force is acting opposite to $u_d$ and will slow down the droplet and eventually reverse its walking direction, and (ii) the magnitude of the horizontal force is decreasing due to the decay of the wave field. In the back-and-forth SGM, the effect of the horizontal force acting opposite to $u_d$ dominates the decay of the wave field and the droplet reverses its walking direction before the wave field has decayed sufficiently. Thus, the droplet is now walking in the opposite direction at a low speed. Eventually the wave field has decayed significantly and the small horizontal force does not have much further impact on the motion of the droplet. 

As $\Delta \phi(t)$ continues to evolve, the droplet transitions back to a $(1,2,1)^{\text{L}}$ mode again and it enters the superwalking regime. The droplet begins to build up the wave field and accelerate again. However, since the droplet was already traveling at a low speed in the opposite direction before the acceleration re-started, the droplet continues to accelerate in the opposite direction. This cycle repeats periodically and results in the back-and-forth SGM.

We note that after reversing the walking direction, the droplet does not immediately accelerate even though it has entered the stage where the constant $\Delta\phi_0$ equilibrium solution has a large superwalking speed. By inspecting the plot of vertical velocity of the droplet $v_d$ at impact and lift-off as a function of the evolving phase difference $\Delta\phi(t)$ in Fig.~\ref{Fig: backforth1}(b), we see that during this time, the droplet is following the equilibrium solution branch corresponding to non-walking solution in the superwalking regime and the droplet begins to accelerate only once it has left this solution branch. This may be due to the droplet losing its wave field during the deceleration stage and hence it takes time for the droplet to build up its wave field again and become unstable to the pure bouncing state. 

For a typical droplet size of $R=0.60$\,mm, such a back-and-forth SGM is realized for a range of negative detunings $-1\leq\epsilon\lesssim -0.2\,$Hz. The horizontal walking velocity $u_d$ and the vertical velocity $v_d$ at impact and lift-off for different detunings are shown in Figs.~\ref{Fig: sGM diff detun}(a) and (c) respectively. We find that as the magnitude of detuning gets smaller in the range $-1\leq\epsilon\lesssim -0.2$\,Hz, the droplet follows the constant phase difference superwalking branch more closely during the deceleration stage.

For small negative detunings $(-0.2\lesssim\epsilon< 0\,\text{Hz})$, the SGM is qualitatively different and we see a mixture of back-and-forth, forth-and-forth and irregular SGM. These are explored in detail in Sec.~\ref{small detun}. Here we note that in the back-and-forth SGM in the small negative detuning regime, see for example the $\epsilon=-0.07$\,Hz curves in Figs.~\ref{Fig: sGM diff detun}(a) and (c), we find that in addition to the droplet following the constant phase difference superwalking branch more closely during deceleration, the droplet only starts accelerating near a phase difference of $\Delta\phi=135^{\circ}$. This also corresponds to a sharp transition from the non-walking solution branch to the superwalking solution branch in Fig.~\ref{Fig: sGM diff detun}(c). 

\section{Forth-and-forth SGM}\label{FF}

\begin{figure}
\centering
\includegraphics[width=\columnwidth]{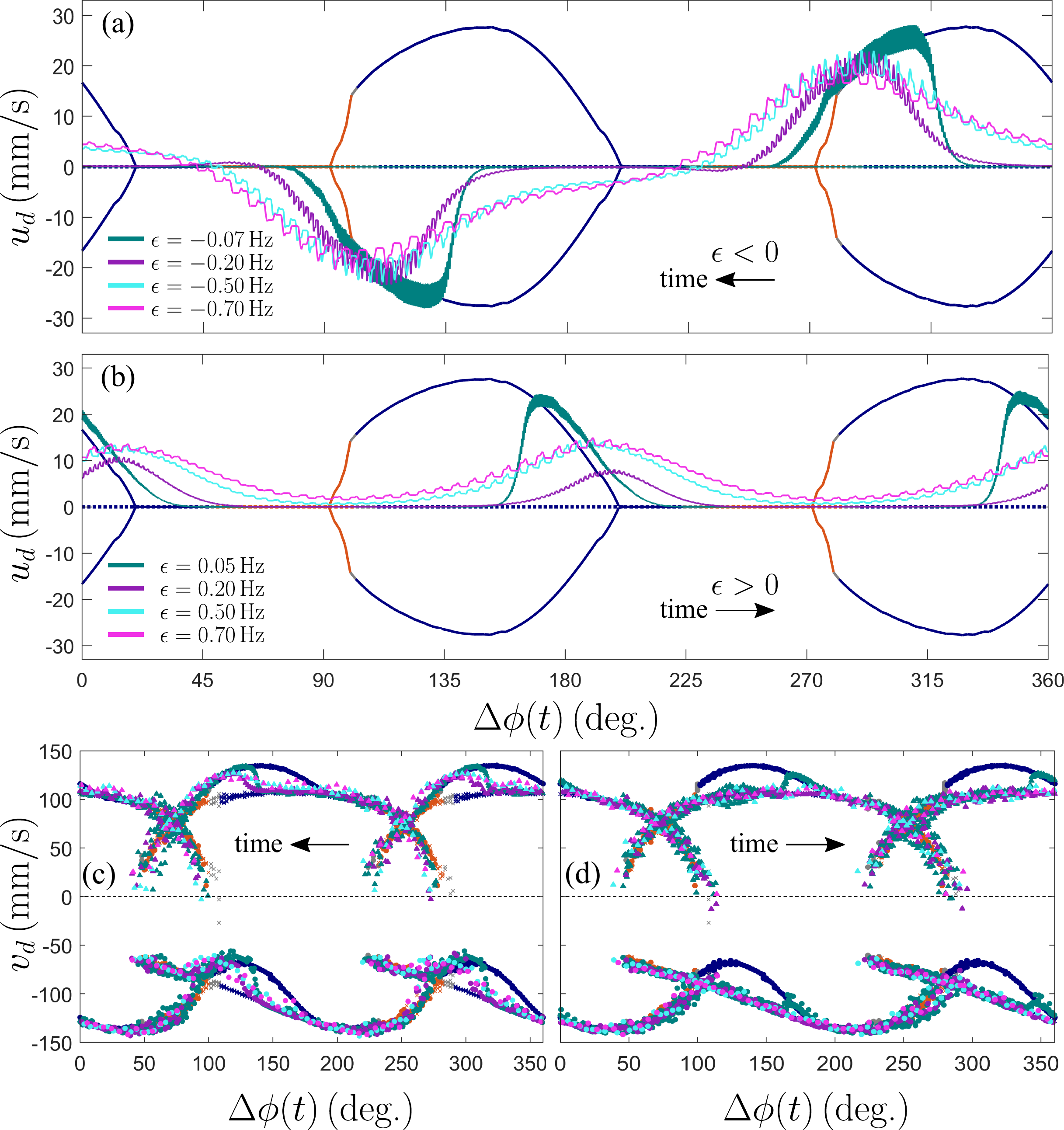}
\caption{Back-and-forth and forth-and-forth SGM for different detunings. Horizontal walking speed $u_d$ as a function of the evolving phase difference $\Delta\phi(t)$ for one cycle of the SGM showing (a) back-and-forth SGM at different negative detunings and (b) forth-and-forth SGM at different positive detunings. Panels (c) and (d) show the vertical velocity of the droplet $v_d$ at impact (lower frame) and lift-off (upper frame) from the back-and-forth and the forth-and-forth SGM corresponding to panels (a) and (b) respectively. The constant phase difference superwalker simulation results are shown as multi-colored curves in the background in these panels.}
\label{Fig: sGM diff detun}
\end{figure}

\begin{figure*}
\centering
\includegraphics[width=2\columnwidth]{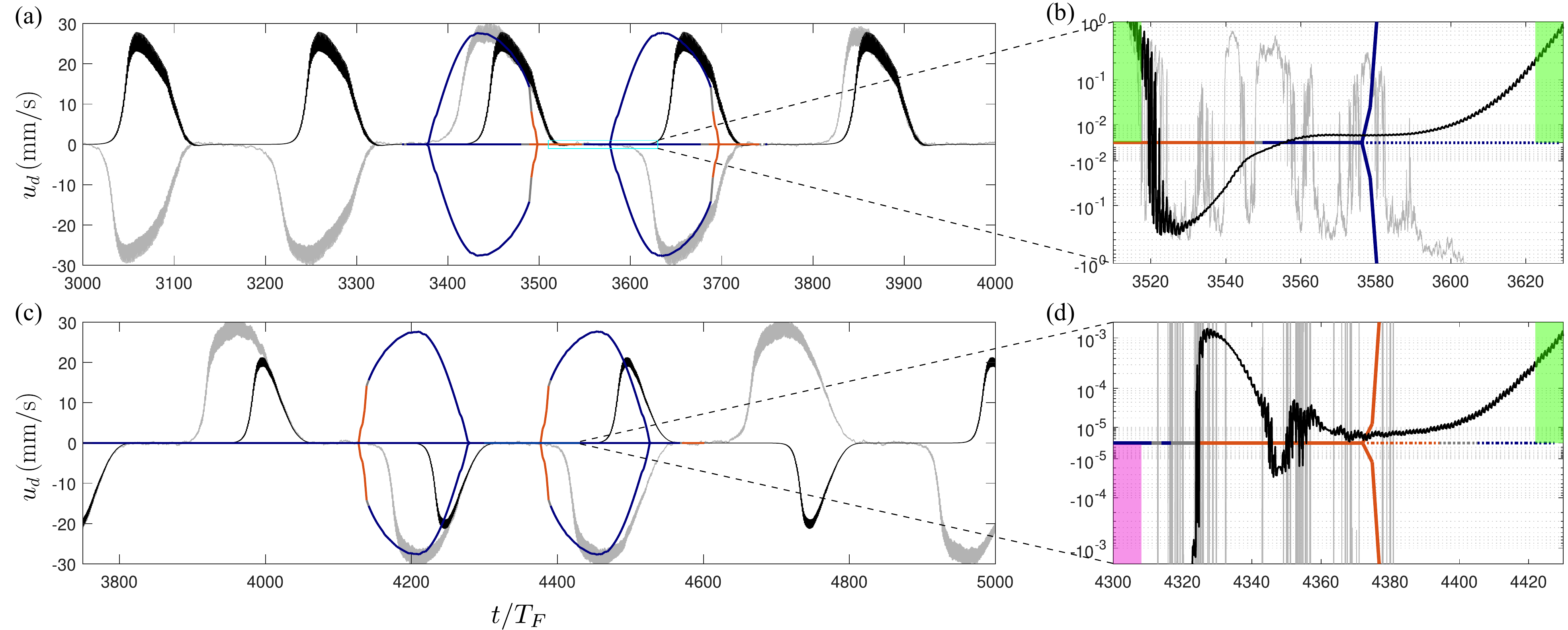}
\caption{Back-and-forth and forth-and-forth SGM for small detunings for a droplet of radius $R=0.60$\,mm. (a) Walking velocity $u_d$ as a function of time for $\epsilon=-0.10$\,Hz showing forth-and-forth SGM without noise (black) and SGM with erratic direction reversals due to added noise (gray). Panel (b) shows a zoomed in section of panel (a) with the vertical axis stretched and the green background highlighting that the droplet continues to walk in the same direction as the previous cycle when no noise is added. (c) Walking velocity $u_d$ as a function of time for $\epsilon=0.08$\,Hz showing back-and-forth SGM without noise (black) and SGM with erratic direction reversals due to added noise (gray). Panel (d) shows a zoomed in section of panel (c) with the vertical axis stretched and the pink and green regions highlighting that the droplet reverses its walking direction compared to the previous cycle when no noise is added. The multi-colored curves represent the corresponding results for constant phase difference simulations.}
\label{Fig: nowalking1}
\end{figure*}

For relatively large droplets and typically positive detunings in Fig.~\ref{Fig: stopgo_PS}, we find that in the SGM, the droplet maintains its walking direction after every half SGM cycle resulting in a forth-and-forth SGM. One cycle of a typical forth-and-forth SGM for $R=0.60$\,mm and $\epsilon=0.5$\,Hz is shown in Fig.~\ref{Fig: forthforth1} (see also Supplementary Material~\citep{supplement} for a video). During the acceleration stage of this SGM, we also find that the droplet bounces in a $(1,2,1)^{\text{L}}$ mode. However, this $(1,2,1)^{\text{L}}$ is qualitatively different from the one observed for back-and-forth SGM, see Fig.~\ref{Fig: backforth1}(c). Here the droplet's vertical motion has three turning points when the droplet is in contact with the bath as compared to a single turning point for the $(1,2,1)^{\text{L}}$ mode in back-and-forth SGM. This difference can be attributed to the fact that in this SGM, as shown in Fig.~\ref{Fig: forthforth1}(b), the vertical velocity of the droplet is unable to completely leave the non-walking branch and transition to the superwalking branch where a $(1,2,1)^{\text{L}}$ mode with only a single turning point during contact would be observed. Instead, the droplet's vertical velocity during the acceleration stage oscillates about the non-walking branch where a qualitatively different $(1,2,1)^{\text{L}}$ mode is realized. However, similar to the back-and-forth SGM, the droplet receives two impulses in the direction of motion and one impulse in the direction opposite to its motion such that the overall horizontal force acting on the droplet is in the direction of its motion allowing it to accelerate. Moreover, during acceleration, the instantaneous velocity of the droplet is smaller than the steady superwalking velocity at constant phase differences.

Once the droplet's instantaneous speed exceeds the corresponding equilibrium speed for constant phase differences, the droplet starts to decelerate. In contrast to the back-and-forth SGM, here we find that the droplet's wave field decays significantly during the initial stages of the deceleration period in the $(1,2,1)^{\text{L}}$ mode. Hence, the horizontal component of the force arising from the gradient of the underlying wave does not a have a significant impact on the droplet. At this time the droplet also transitions to a $(1,2,2)$ mode in which the droplet is unable to build its wave field. Since the droplet is already near the peak velocity at the start of the deceleration stage, it experiences large drag forces that eventually slow it down. But unlike the back-and-forth SGM, the droplet's walking direction does not get reversed as the drag force can only slow the droplet down and not reverse its walking direction. Hence, the droplet continues to walk with a decaying speed in the same direction as the $\Delta\phi(t)$ passes through the bouncing regime. Eventually, the droplet enters the phase differences for the superwalking regime where the droplet's steady superwalking velocity exceeds its instantaneous velocity and the droplet also transitions to a $(1,2,1)^{\text{L}}$ mode, where it can build up its wave field and start to accelerate again. Since the droplet already had an initial perturbation in the direction of motion before it entered the acceleration stage, the motion continues in the same direction resulting in forth-and-forth SGM. 

We find that after the velocity of the droplet has decayed significantly and is very small, the droplet does not immediately accelerate even though it has reached the phase differences where the equilibrium solution has a large superwalking speed. Here we find that, in comparison to the back-and-forth SGM, the droplet takes a longer time to build up the wave field and accelerate and thus the peak speed it can reach before it starts to decelerate is smaller compared to back-and-forth SGM. This again can be attributed to the droplet never completely transiting to the superwalking branch in Fig.~\ref{Fig: backforth1}(b), and thus being unable to accelerate efficiently.

For a typical droplet size of $R=0.60$\,mm, such a forth-and-forth SGM is realized for a range of positive detunings $0.2\lesssim\epsilon\leq0.7$\,Hz. The horizontal walking velocity $u_d$ and the vertical velocity $v_d$ at impact and lift-off for different positive detunings is shown in Figs.~\ref{Fig: sGM diff detun}(b) and (d) respectively. We find that as the magnitude of detuning increases from $0.2$\,Hz to $0.7$\,Hz, the droplet is able to achieve slightly higher peak speed during the `go' phase. This is also correlated with the increase in the magnitude of oscillations around the non-walking branch in the superwalking regime as the detuning increases. 

For small positive detunings $(0<\epsilon\lesssim 0.2\,\text{Hz})$, the SGM is again qualitatively different and we see a mixture of back-and-forth, forth-and-forth and irregular SGM. It is explored in detail in Sec.~\ref{small detun}. Here we note that in the forth-and-forth motion that is realized for small positive detunings (see $\epsilon=0.05$\,Hz curves in Figs.~\ref{Fig: sGM diff detun}(b) and (d)), we find that the droplet is able to completely leave the non-walking branch and transition to the superwalking branch and thus achieve a significantly higher superwalking speed in the forth-and-forth SGM compared to the ones described above. 

\begin{figure}
\centering
\includegraphics[width=\columnwidth]{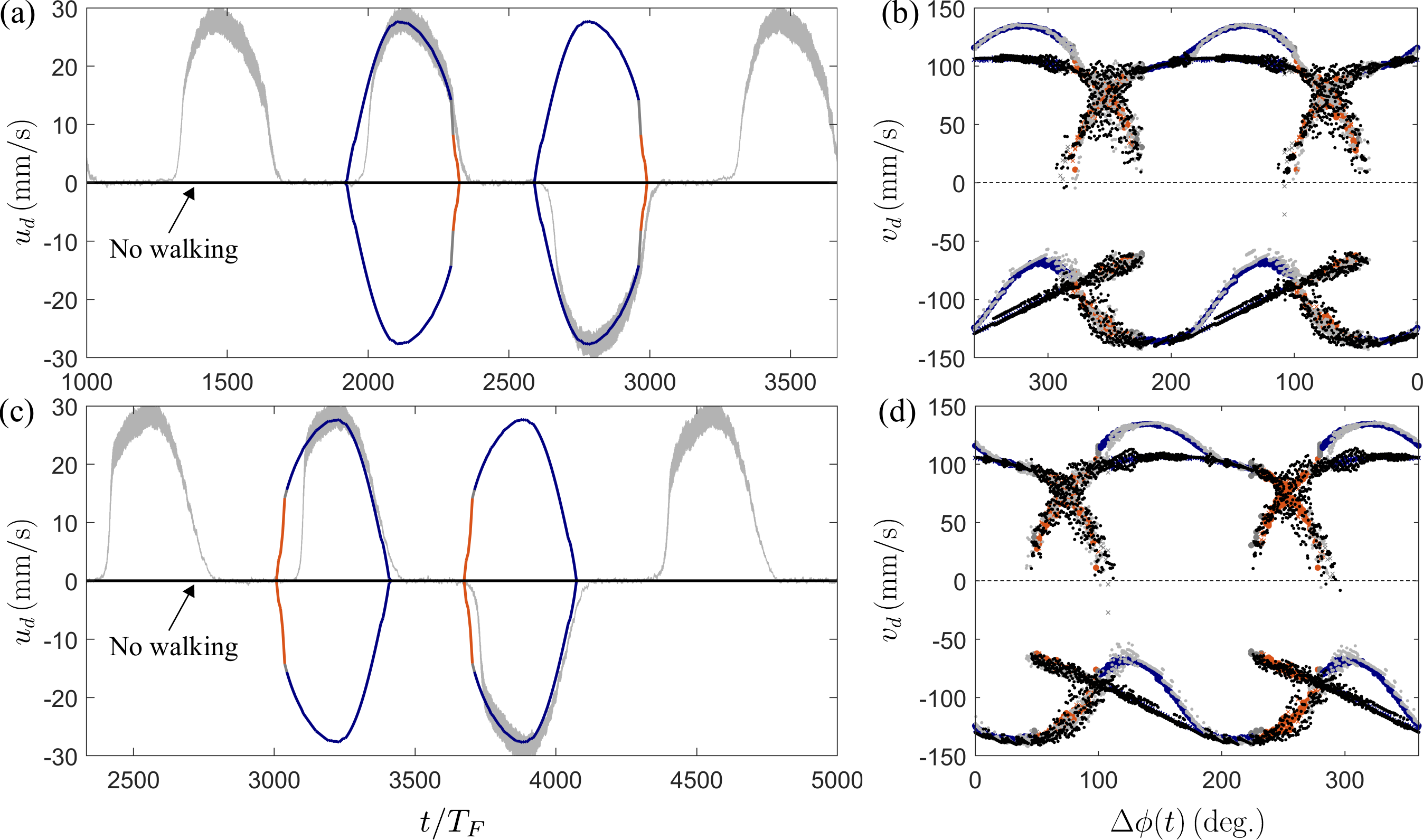}
\caption{No walking for very small detunings $(|\epsilon|<0.04)$ and the effect of added noise for a droplet of radius $R=0.60$\,mm. Panels (a) and (c) show the walking velocity $u_d$ as a function of time without noise (black) and with noise (gray) for $\epsilon=-0.03$\,Hz and $\epsilon=0.03$\,Hz respectively. Panels (b) and (d) shows the vertical velocity of the droplet $v_d$ at impact and lift-off without noise (black) and with noise (gray) corresponding to panels (a) and (c) respectively. The constant phase superwalker simulations results are shown as multi-colored curves in these panels.}
\label{Fig: nowalking2}
\end{figure}

\section{SGM for small detuning}\label{small detun}

For both positive and negative detunings that are relatively small in magnitude ($-0.2\lesssim\epsilon\lesssim-0.05$\,Hz and $0.05\lesssim \epsilon \lesssim 0.2$\,Hz), we observe a mixture of back-and-forth, forth-and-forth and irregular SGM for a typical droplet size of $R=0.60$\,mm. However, we find that these SGMs are qualitatively different from the ones that emerge at larger magnitudes of positive and negative detunings. 

In the irregular SGM that emerges in this regime for small detunings, the droplet travels the same distance every half SGM cycle while the reversals of direction are erratic. In contrast, the irregular SGM that arises at large detunings in the parameter space for relatively small droplets, the droplet traverses unequal path lengths as well as performing erratic direction reversals every half cycle of the SGM.

The back-and-forth SGM discussed in Sec.~\ref{BF} reverses its walking direction only once every half cycle and then continues moving in the opposite direction while the forth-and-forth SGM discussed in Sec.~\ref{FF} never reverses its walking direction. As shown in Figs.~\ref{Fig: nowalking1}(a-d), for both back-and-forth and forth-and-forth SGM at small detunings, we find that the droplet reverses its direction multiple times when traversing the bouncing regime. For this regime of SGM, when the droplet first reaches the phase differences corresponding to the bouncing regime, the droplet decelerates. Since the detuning is small, the bouncing regime is traversed slowly and the droplet first overshoots the zero velocity while decelerating and reverses it walking direction. But the droplet cannot accelerate yet in the opposite direction since it is still in the bouncing regime, and therefore the droplet approaches the zero velocity solution again and overshoots, resulting in a second reversal of the walking direction. These reversals of direction continue until the droplet reaches the start of the superwalking regime where it can begin to build up the wave field again and then accelerate. Whichever direction the droplet happens to move when it encounters the start of the superwalking regime, it continues to walk and accelerate in that direction. If the number of zero velocity crossings are even, then a forth-and-forth SGM emerges and if the number of crossings are odd, then back-and-forth SGM is realized. 

By performing simulations with added noise in the form of a horizontal random force sampled from a uniform distribution in the range $F_{noise}=[-10^{-7},10^{-7}]$\,N, much smaller in magnitude than the typical forces experienced by the droplet from the underlying wave, we find that we are able to destroy these correlated back-and-forth and forth-and-forth motions at small detunings (see gray curves in Fig.~\ref{Fig: nowalking1}). In these simulations, an SGM with irregular direction reversal emerges due to the random external noise added to the system. In typical experiments where small noise sources are to be expected, back-and-forth and forth-and-forth SGM may not be realized at small detunings. We also note that for these simulations with added noise, the droplet's vertical velocity is able to leave the non-walking branch more easily and following the superwalking branch closely.

For very small detunings $|\epsilon|\leq 0.04$\,Hz, we find that the SGM ceases to exist and the droplet remains stationary. Fig.~\ref{Fig: nowalking2}(a-d) shows the walking velocity $u_d$ as a function of time and the vertical velocity of the droplet $v_d$ at impact and lift-off as a function of the evolving phase difference. Here we see that the droplet always follows the non-walking branch for all phase differences, which means that the droplet is not able to transition to the superwalking branch and build up the wave field it needs to propel itself. However, we find that adding noise in the horizontal dynamics, similar to Fig.~\ref{Fig: nowalking1}, destroys this fragile non-walking state and we again observe SGM, with random switches in the walking direction resulting from the noise.

\section{Conclusions}\label{DC}


We have numerically investigated a new type of intermittent locomotion, the stop-and-go motion (SGM) of superwalking droplets, that emerges when a liquid bath is driven simultaneously at frequencies $f$ and $f/2$ with an evolving phase difference $\Delta\phi(t)=2\pi \epsilon t$. The SGM is a complex nonlinear phenomenon with multiple timescales coming into play such as the bouncing time scale of the droplet, the memory time scale associated with decay of Faraday waves, the even longer time scale introduced by the detuning and the time scale of the inertial response of the droplet. We have observed three qualitatively different kinds of SGM in simulations: back-and-forth, forth-and-forth and irregular. We note that motion similar to irregular SGM reported here has also been observed in experiments with two frequency driving with frequencies $80$\,Hz and $64$\,Hz~\citep{PhysRevE.94.053112}. In preliminary experiments with the setup described in \citet{superwalker}, we have frequently observed back-and-forth SGM. We have also observed irregular SGM often, and less frequently what appears to be forth-and-forth SGM. However, it is difficult to differentiate between a forth-and-forth and irregular SGM in such an experimental setup as the horizontal extent of the domain is limited and a droplet undergoing a forth-and-forth SGM would encounter the boundary after only a few cycles of the SGM making it difficult to conclude whether it is truly forth-and-forth motion. A detailed experimental study of SGM is needed to reveal the different types of SGM that may be realized in laboratory.  An annular domain might be appropriate to study forth-and-forth SGM experimentally. 

Investigating the back-and-forth SGM for negative detunings that are relatively large in magnitude reveals the mechanism for the switch in the walking direction that occurs after every half SGM cycle. In this SGM, the droplet receives persistent impulses in the direction opposite to its motion during its deceleration stage. These persistent impulses eventually slow down the droplet and reverse its walking direction before the underlying wave field decays significantly. Hence, the droplet already has an initial perturbation in the direction opposite to the direction of motion when it encounters the superwalking regime again and it is able to accelerate in the opposite direction for the next cycle of SGM. We note that reversals of motion have also been demonstrated in the system of single-frequency driven walkers where a pulse in the driving signal was engineered to abruptly introduce a $\pi$ shift in the bouncing phase of the droplet, which results in the droplet retracing its previous trajectory~\citep{perrard2016}. Here we observe this behavior arising as an emergent phenomenon.

Investigating the forth-and-forth SGM for large positive detunings reveals the mechanism for the droplet to maintain its walking direction after every half SGM cycle. In this SGM, the droplet's wave field decays significantly during the initial stages of the deceleration period in the bouncing regime and hence for the remainder of the deceleration stage, only the drag force is acting on the droplet. The drag slows down the droplet but cannot reverse the walking direction so when the droplet encounters the superwalking regime again, it accelerates in the same direction. It is worth noting that in both the back-and-forth and forth-and-forth SGM, the periodic reversal or maintenance of the walking direction is not directly due to the memory of the waves that are generated by the droplet since the wave field decays significantly during each cycle for both kinds of SGM.

For small positive and negative values of detunings, we found a mixture of back-and-forth, forth-and-forth and irregular SGM. The back-and-forth and the forth-and-forth here are qualitatively different from the ones at large magnitude detunings. Here we find that the droplet switches its direction multiple times while traversing across the bouncing regime. Moreover, these SGM also tend to be less robust and adding a small noise force destroy the back-and-forth and forth-and-forth SGM in this regime. 

In this work, we considered SGM that is inspired by the slight detuning of the two driving frequencies in superwalker experiments that results in the phase difference varying linearly in time. Using the back-and-forth and forth-and-forth SGM as building blocks, one might be able to control and program the droplet's motion by engineering any slowly varying time-dependent functions $\Delta\phi(t)$. Moreover, walkers and superwalkers are known to exhibit rich multi-droplet behaviors. It would be interesting to investigate such multi-droplet interactions in the SGM regime where the continuously changing phase difference will result in periodically varying inter-droplet interactions that may give rise to novel collective behaviors.

\acknowledgements
We acknowledge financial support from an Australian Government Research
Training Program (RTP) Scholarship (R.V.) and the Australian Research Council via the Future Fellowship Project No.\ FT180100020 (T.S.).

\bibliography{SGM}

\end{document}